\documentclass[aps,pre,preprint,superscriptaddress]{revtex4-1}

\usepackage{graphicx}
\usepackage{epstopdf}
\usepackage{caption,booktabs}
\usepackage{subcaption}
\begin{document}


\title{Variation of the critical percolation threshold in the Achlioptas processes}


\author{Paraskevas Giazitzidis}
\affiliation{Department of Physics, University of Thessaloniki, 54124 Thessaloniki, Greece}
\author{Isak Avramov}
\affiliation{Institute for Physical Chemistry, Bulgarian Academy of Sciences, 1113 Sofia, Bulgaria}
\author{Panos Argyrakis}
\affiliation{Department of Physics, University of Thessaloniki, 54124 Thessaloniki, Greece}


\date{\today}

\begin{abstract}
We investigate variations of the well-known Achlioptas percolation problem, which uses the method of probing sites when building up a lattice system, or probing links when building a network, ultimately resulting in the delay of the appearance of the critical behavior. In the first variation we use two-dimensional lattices, and we apply reverse rules of the Achlioptas model, thus resulting in a speed-up rather than delay of criticality. In a second variation we apply an attractive (and repulsive) rule when building up the lattice, so that newly added sites are either attracted or repelled by the already existing clusters. All these variations result in different values of the percolation threshold, which are herewith reported. Finally, we find that all new models belong to the same universality class as classical percolation.  
\end{abstract}

\pacs{82.40.Ck, 89.75.Fb, 64.60.ah, 61.43.Bn, 05.70.Fh}

\maketitle


\section{Introduction}

The percolation phase transition \cite{Stauffer} has traditionally attracted the interest not only of physicists but of scientists in practically all fields over the last five decades due to the fact that it is a paradigmatic continuous phase transition. More recently, new models have appeared \cite{achlioptas2009} which vary the way by which a lattice is filled up, thus producing a different critical threshold value, and very different characteristics of the phase transition, including the question if the well known transition is continuous or discontinuous. The delay of criticality in the percolation problem introduced in \cite{achlioptas2009} has certainly attracted considerable interest in recent years because it gives one a method to vary the exact location of the critical point, almost at will, by appropriately handling the conditions by which the system is being built up or prepared \cite{bohman,Adler1991,reviewziff}. The initial idea to the direction of discontinuous phase transition included a 2-particle probe method, which caused the critical point to be considerably delayed. Thus, for a square 2D lattice the critical point moved from $p_c=0.593$ to $p_c=0.755$. A similar behavior can be seen in different geometries, as well as in networks, where the critical point refers now to the creation of the largest spanning cluster. Recently, we showed \cite{paris2013} that by using a different number of probe sites one can further vary at will the location of the critical point, showing a well-behaved monotonic behavior of the location of the critical point as a function of the number of probe sites. Further variations include the use of the product-rule vs the sum-rule and have implemented either in lattices or in networks \cite{review2014}, in which case one uses either the product of the sizes of the clusters to be connected or the sum of the same sizes, respectively. While qualitatively the same result is obtained, the exact location  of the critical point depends on the criterion used (product, sum or other similar rule). 

In this work we present two new models. In the first model the critical point appears earlier than the classical critical point ($p'_c <p_c^{classical}$). To do this we use the same idea as the original Achlioptas probe method, but instead of choosing the site that results in the smaller product or sum of the joining cluster we now choose exactly the opposite, i.e. we choose to keep the probe site which results in the largest cluster and we discard the other probe site. Obviously, this will result  in a speed-up of the critical point, i.e. the largest percolating cluster will now appear earlier than the conventional case. Indeed we find that instead of $p_c=0.593$ we now have $p_c=0.531$. Thus, the speed-up of the critical point gives the complimentary case of the celebrated Achlioptas model. 

In a second model we introduce a different method of probing sites  in a percolating system. Initially we choose at random a site in the system and we decide to occupy it only if it has at least one occupied nearest neighbor. If the randomly chosen site has no neighbors at all then we decide not to occupy it. In this case we randomly choose another site and we occupy it regardless of the number of its neighbors. This procedure promotes the attraction between nearest sites and for this reason large clusters are merging faster than in the classical percolation case. This attracting process results again in the early emergence of the critical point. We find this speed-up in criticality to produce a critical point of $p_c=0.562$ instead of $p_c=0.593$. We also evaluated the opposite process by promoting the isolated sites and suppress the emergence of large clusters. This time sites with no occupied neighbor are preferred, and as expected a delay to criticality occurs. The critical point now rises to $p_c=0.610$.       

\section{Model description}

We investigate four variations of the well known Achlioptas processes \cite{achlioptas2009}. In the original model one fills the system (lattice or network) by probing at random two candidate sites (or nodes) to be occupied. The one that minimizes the product of the sizes of the clusters that this site is about to connect is maintained, while the other one is removed. The details for candidate sites maybe different according to the system used. For example, in site percolation in a two-dimensional (2D) square lattice there is a maximum of four possible clusters that can be merged, while in the original Achlioptas processes the newly added bond may connect only two clusters to form a larger one. However, such details do not affect the overall system behavior. We have reproduced the data for site percolation product-rule and sum-rule and our results are in excellent agreement with previous publications\cite{bastas2011,choi2012}. Here we extend both these  models by promoting the creation of larger clusters, instead of smaller ones. In order to do this we first choose two candidate sites. We calculate the product of the sizes of the clusters that are to be merged for each candidate site separately. Then we keep the site with the larger of the two products, while we discard the other one. This results in the critical point appearing earlier than in normal percolation. Thus, in addition to the delay of criticality that was suggested by the Achlioptas models, one may now speed-up the appearance of the critical point.    

Furthermore, we now introduce two new models, which are based on filling the lattice by probing the local environment of the site to be added, one using an attractive algorithm, and one using a repulsive one. In the attraction model we start initially with an empty 2D square lattice of linear size $L=1000$. We start by probing one site of the lattice at random and we occupy it only if this site has at least one neighbor occupied. We consider that each site has four nearest neighbors. If the chosen site has no nearest neighbors occupied, we then choose at random another site of the lattice without investigating if there are any occupied neighbors (random site percolation) and we occupy it. We continue by probing a second site, and so on, and we repeat the same process as previously. In the repulsion model we start again with an empty 2D square lattice of the same linear size $L=1000$ and we probe a site at random. We then investigate the four neighbor sites of the candidate site and we occupy it only if there are no occupied neighbors at all. On the other hand, if there is at least one neighbor site that is occupied, we choose another site at random and we occupy it. For both the attraction model and repulsion model we continue until the lattice is fully occupied.

These two models, the attraction and the repulsion model, are expected to significantly change the location of the critical point (as discussed in the section of Results). The location of the critical point can conceivably be further changed if one is increasing the number of attempts, for both the attraction and the repulsion models, which is expected to increase the level of speed-up or delay of the critical point, respectively. To investigate this we consider the same 2D square lattice of linear size $L=1000$, but this time we introduce a parameter $k$ which gives the number of attempts when probing the lattice sites. Thus, for $k=0$ this corresponds to conventional random site percolation. $k=1$ gives the model explained in the previous paragraph, where we have only one attempt to search for occupied nearest neighbors before we probe a site of the lattice at random. $k=2$ results to two independent attempts. More specifically, in the attraction model we start again with an empty lattice and we probe at random a site to be occupied. If this chosen site has no nearest neighbors occupied, we probe another site at random and we investigate again for occupied nearest neighbors. That was the second attempt $(k=2)$. If again no nearest neighbors are occupied we occupy a random site in the lattice without checking its neighbors. It is important to mention that if we find a nearest neighbor occupied during the attempts, the process stops and a new Monte Carlo step starts from the beginning. This procedure can be extended to larger $k$ values. We report simulations with different $k$ values for both the attraction and the repulsion models.  

\section{Results}

We monitor the percolation strength $P_{max}$, which gives the probability of a given occupied site to belong to the largest percolating cluster. $P_{max}$ is in the range $0 < P_{max} < 1$ and it represents that part of the system that is been occupied by the percolating cluster.  
Eq. \ref{eq1} shows $P_{max}$ as a function of the density of occupied sites $p$.

\begin{equation}
P_{max}= \frac{S_1}{p L^2}
\label{eq1}
\end{equation}

where $S_1$ is the size (number of sites) of the largest cluster of the system at density $p$, and $L^2$ is the total number of lattice sites.

\begin{figure}
 \linespread{1}
 \includegraphics[width=\linewidth]{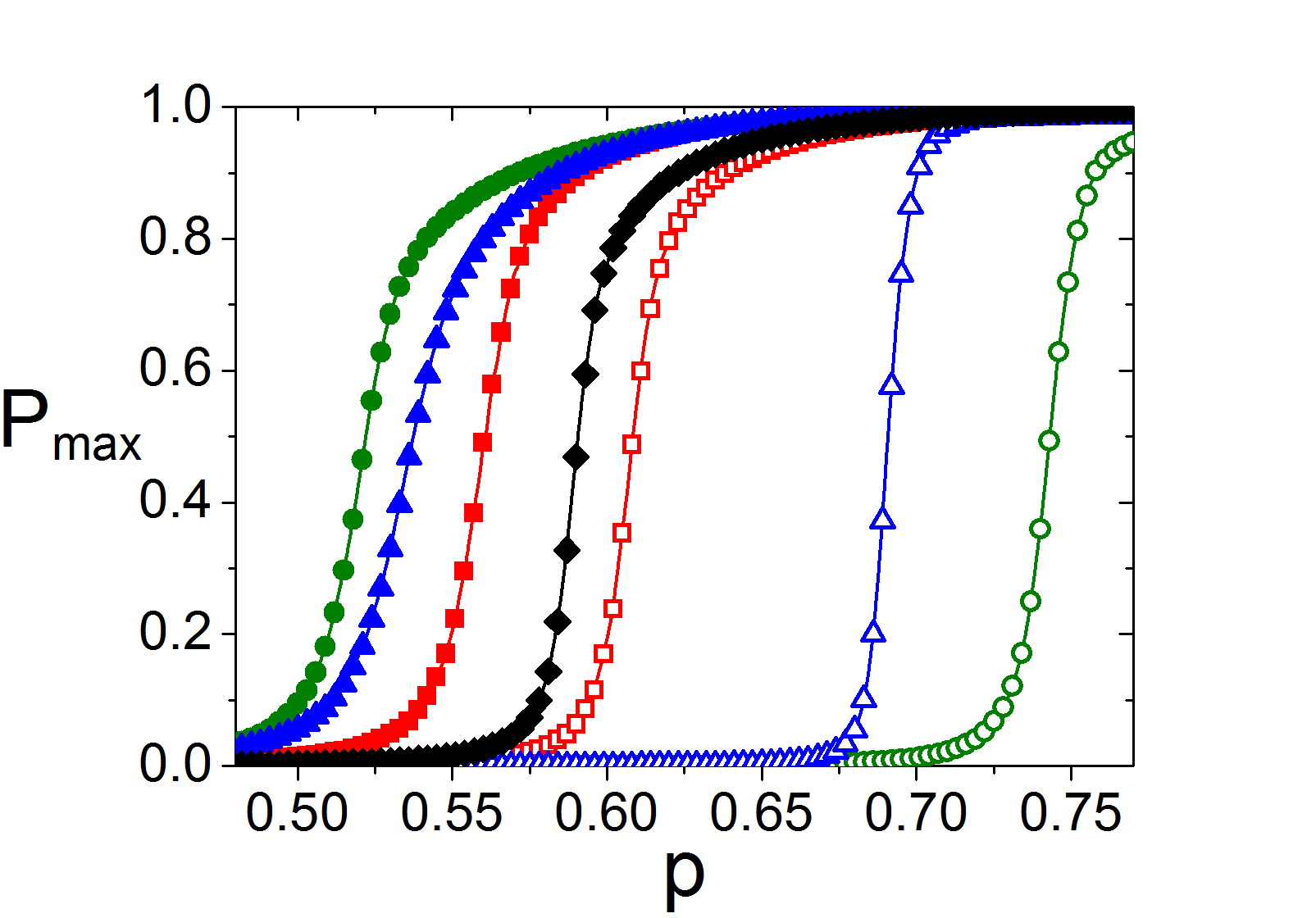}
 \caption{\label{fig1} Plot of percolation strength ($P_{max}$) as a function of the density of occupied sites $p$ for a 2D square lattice of linear size $L=1000$. Classical site percolation is indicated with black full diamonds. The attraction model($k=1$) is indicated with red full squares and the repulsion model($k=1$) with the empty ones. Sum-rule is indicated with blue triangles. Empty ones are for the delay of criticality, while the full triangles are for the speed-up version. Product-rule is indicated with green circles. Empty circles are for original product rule for the delay of criticality, while full circles are again for the speed-up version.  The lines are optical guides.}
 \end{figure}

\begin{figure}
\linespread{1}
  \includegraphics[width=\linewidth]{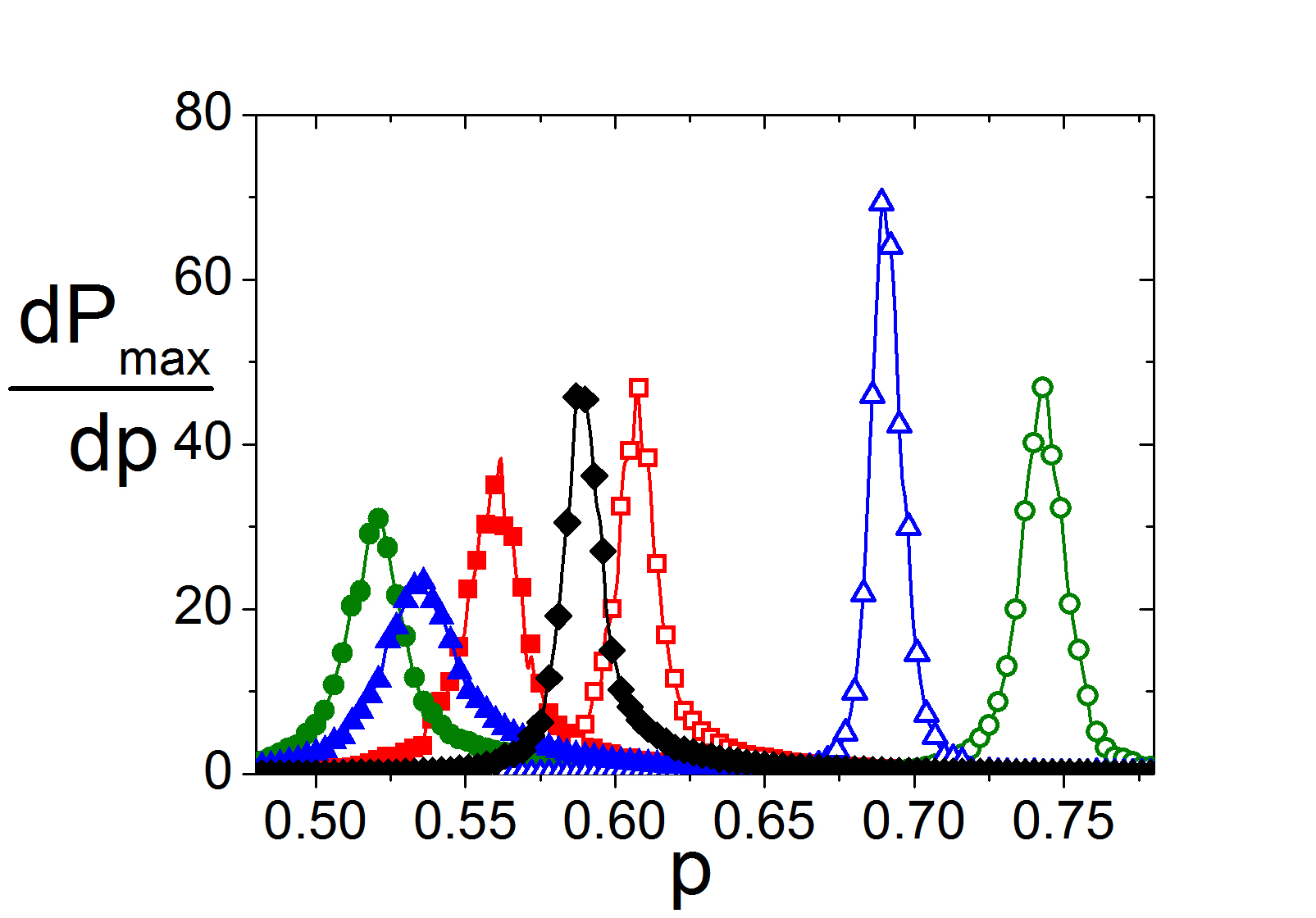}
  \caption{\label{fig2} Plot of the first derivative ($\frac{P_{max}}{dp}$) of percolation strength as a function of the density of occupied sites $p$ for a 2D square lattice of linear size $L=1000$. Classical site percolation is indicated with black full diamonds. The attraction model($k=1$) is indicated with red full squares and the repulsion model($k=1$) with the empty ones. Sum-rule is indicated with blue triangles. Empty ones are for the delay of criticality, while the full triangles are for the speed-up version. Product-rule is indicated with green circles. Empty circles are for original product rule for the delay of criticality, while full circles are again for the speed-up version.  The lines are optical guides.}
  \end{figure}

In Fig. \ref{fig1} we give $P_{max}$ for seven different models. The classical site percolation gives $p_c = 0.593$ (full diamonds). The Achlioptas product (PR) and sum-rule (SR) for the delay  of criticality are shown with $p_c = 0.755$ (open circles) for PR, $p_c=0.694$ (open triangles) for SR. The equivalent Achlioptas processes for the early emergence of criticality now produce  $p_c = 0.531$ (full circles) for PR, $p_c = 0.543$ (full triangles) for SR. The critical threshold for the attraction model is $p_c = 0.562$ (red full squares) and for the repulsive model is $p_c = 0.610$ (red empty squares). While the values of these critical points can be deduced from Fig. \ref{fig1} as the mid-point of the sudden transition, it is usually better for higher accuracy to calculate the first derivative of $P_{max}$, as this shows the transition in a sharp fashion. We, thus, calculate $\frac{dP_{max}}{dp}$ for all seven cases, and we show the results in Fig. \ref{fig2}. We observe that the peak of the curvature of the first derivative gives the critical density $p_c$ for each model. The values of $p_c$ for product-rule and sum-rule as well as for the classical percolation are in excellent agreement with previous publications \cite{ziff2000,bastas2011,choi2012}, while values for critical densities for the reverse processes of product and sum-rule are now calculated here.

The Table illustrates the critical threshold for all seven different models that are included in Fig. \ref{fig1}.
To see if these two new processes belong to the same universality class with classical random percolation, we now calculate the universal critical exponent which does not depend on the structural details (topology) of the lattice or on the type of percolation (site, bond). This exponent is the fractal dimension $d_f$. The universality property is a general property of second order phase transitions, where the order parameter (here is the size of the infinite cluster $S_1$) introduces an abrupt increase at the region near the critical point. 
Eq. \ref{eq3} shows the logarithmic relation of $S_1$ to the linear size of the lattice $L$ around the critical point $(p \approx p_c)$. 

\begin{table} 
 \caption{\label{table}  The critical threshold $p_c$ and the fractal dimension critical exponent $d_f$ for all models. The results are for site percolation transition on a 2D square lattice.}
 \begin{ruledtabular} 
  
 \begin{tabular}{l {c}r}

Model           & $p_c$& $d_f$  \\
 \hline
 Classical percolation           & 0.5927 & $1.89 \pm  0.02$ \\
 Attraction model $(k=1)$	&  0.5618 & $1.89 \pm  0.03$   \\
 Repulsion model $(k=1)$ & 0.6100  & $1.88 \pm  0.03$ \\
 Product rule (delay)        & 0.7554 & $1.99 \pm  0.01$ \\
  Product rule (early emergence)       & 0.5315 & $1.87 \pm  0.02$ \\
   Sum rule (delay)        & 0.6942  & $1.99 \pm  0.01$ \\
    Sum rule (early emergence)       & 0.5433  & $1.88 \pm  0.02$ \\
 \end{tabular}
  
 \end{ruledtabular}
 \end{table}

\begin{equation}
S_1(L)_{(p \approx p_c)} \sim L^{d_f}
\label{eq3}
\end{equation}

\begin{figure}
\linespread{1}
  \includegraphics[width=\linewidth]{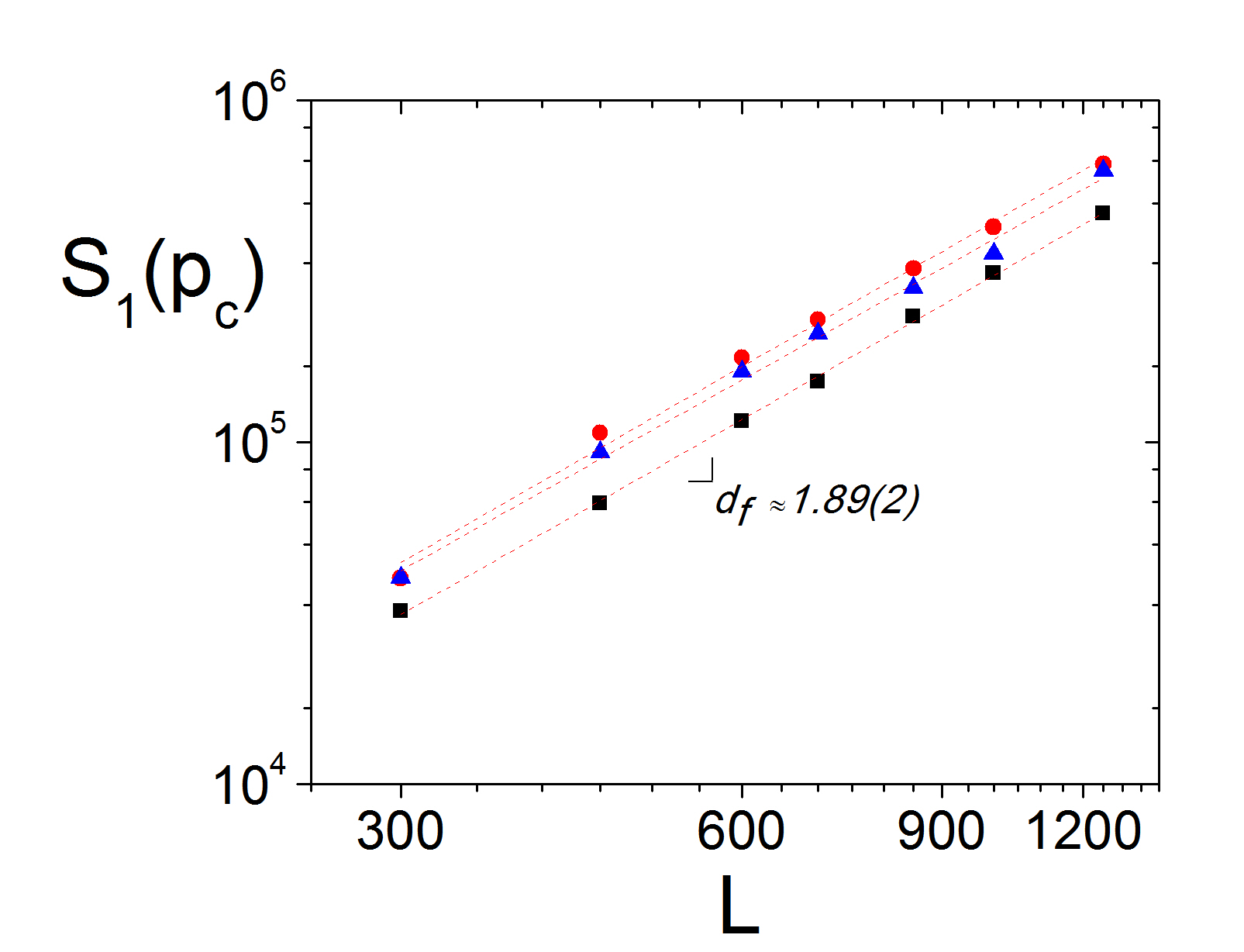}
  \caption{\label{fig3} Finite size scaling of the size of the largest cluster at the critical point $S_1(L)_{(p=p_c)}$ as a function of the linear lattice size $L$ for the attraction (red circles) and the repulsion (blue triangles) models in log-log coordinates. Black squares are for the classical percolation case. The slopes of the straight lines give the fractal dimension $d_f$. All curves have the same $d_f$, meaning that all different models belong to the same universality class.}
  \end{figure}

Fig. \ref{fig3} illustrates the scaling of the size of the largest cluster at the critical point $S_1(p=p_c)$ as a function of $L$. The slope of the straight line gives the fractal dimension $d_f$, which for the classical random percolation is well-known and its value has been calculated $(d_f \approx 1.896)$ \cite{Stauffer}. We observe that the slopes of the three straight lines in Fig. \ref{fig3} have the same value, meaning that all different models examined here belong to the same universality class.

\begin{figure}
\linespread{1}
\centering
\begin{subfigure}{.3\textwidth}
 \centering
  \includegraphics[width=\linewidth]{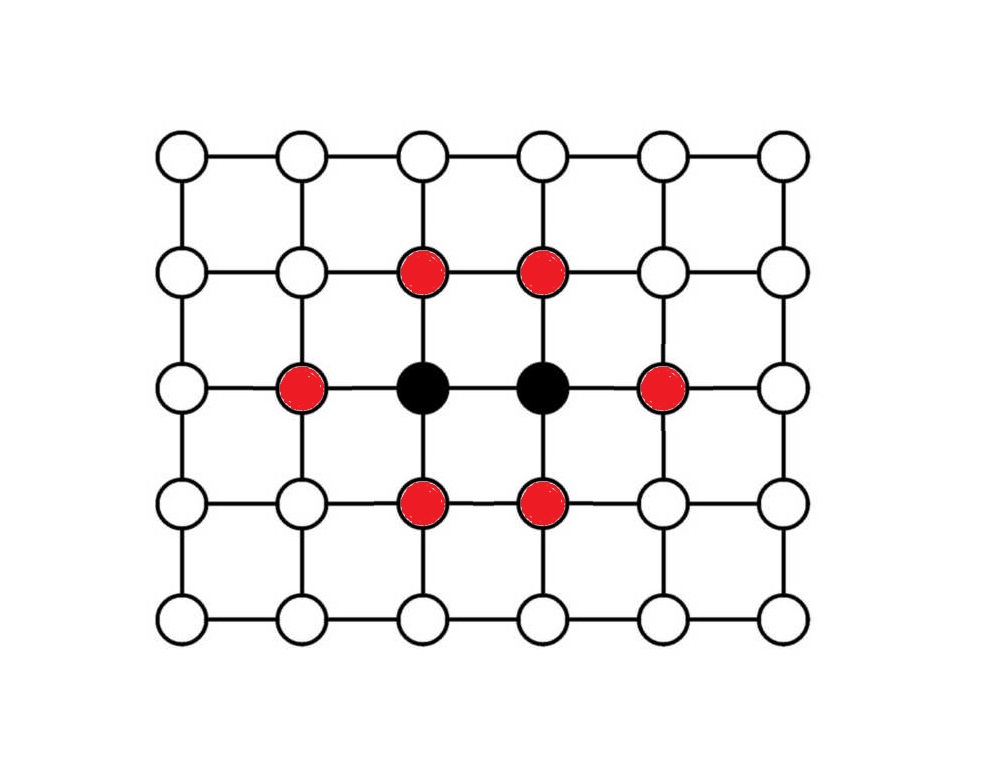}
  \caption{}
 
\end{subfigure}%
\begin{subfigure}{.3\textwidth}
 \centering
  \includegraphics[width=\linewidth]{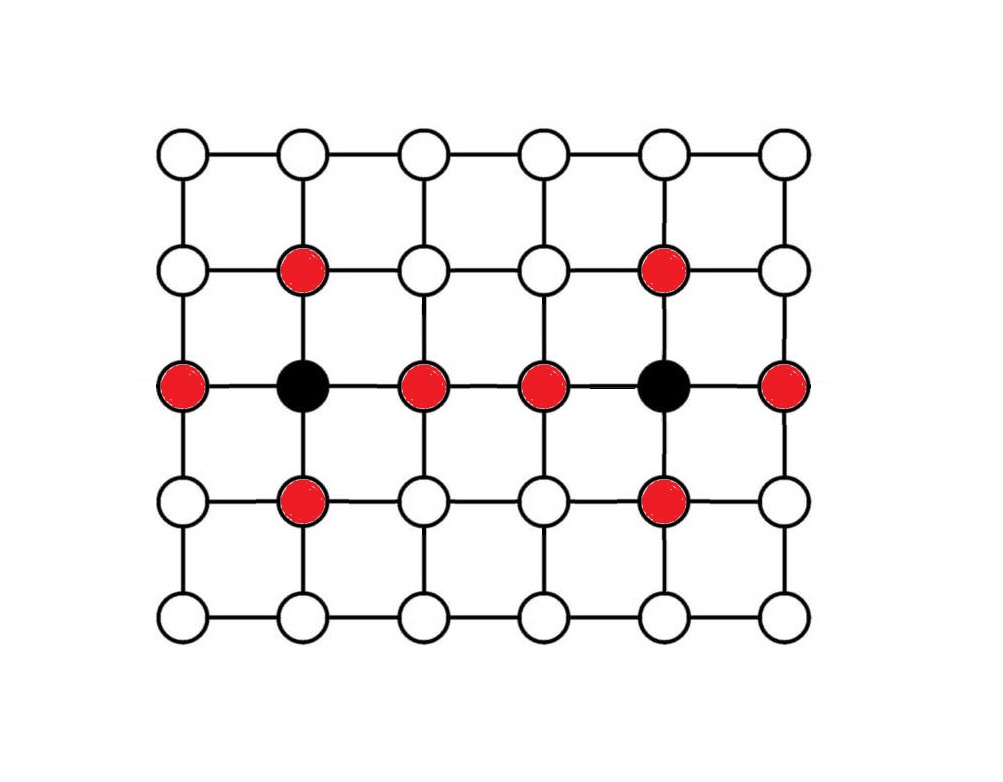}
  \caption{}
  
\end{subfigure}
\caption{The difference in the construction process between attraction model $(a)$ and repulsion model $(b)$. White sites are empty, black sites are occupied, while red ones are those with at least one occupied nearest neighbor. There are more red sites for a fixed number of black at the repulsion model.}
\label{fig4}
\end{figure}

\begin{figure}
\linespread{1}
 \includegraphics[width=\linewidth]{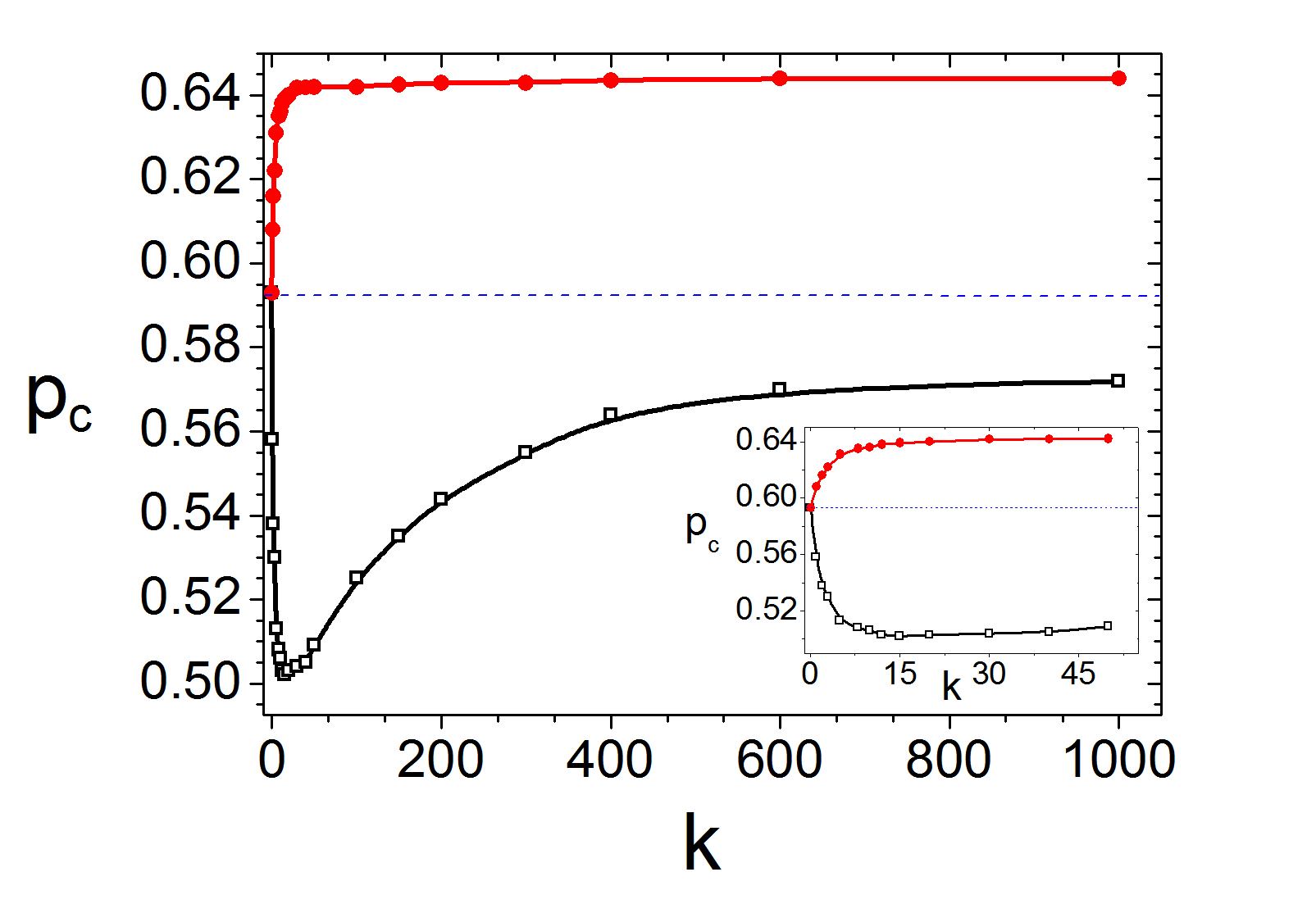}
 \caption{\label{fig5} The percolation threshold of sites $p_c$ vs the number of independent attempts $k$ for attraction and repulsion models. The attraction model is illustrated with empty squares while the repulsion model with full circles. The lines are only optical guides. The blue dashed line indicates $p_c$ for classical percolation. In the inset: Critical density of sites $p_c$ vs number of attempts $k$ to indicate the minimum and the maximum of the two curves. }
 \end{figure}

One can see an interesting observation in Fig. \ref{fig1}. This is the point that the curves for the attraction model and repulsion model are not symmetric around the curve for the classical percolation. This is due to the fact that the rule that we use is not  exactly equivalent for both models. For very low densities ($p < 0.1 $) the majority of the newly added sites are randomly distributed in the system for both models. There are many isolated sites and the majority of the system consists of empty space. In the attraction model the newly chosen site is isolated in most of the times and thus, a new site is chosen to be occupied (random percolation). In the repulsion model again we chose at random to occupy a site. Since it is more probable that there are no occupied nearest neighbors, we occupy this specific site, which at these low values of $p$, occurs most frequently in almost every Monte Carlo step. But again, this procedure is equivalent to random percolation. At higher density values but still lower than the critical, $(0.1 < p < 0.5)$, in the attraction model the system consists of small clusters. In contrary, in the repulsion model the system is sparse (the majority of occupied sites is isolated). As a result of this, when we choose to occupy a site at random, in the attraction model this site is attached to the clusters that already exist in the system, while in the repulsion model, it is more likely for each randomly probed site to have an occupied nearest neighbor, and thus a new site is been chosen at random. This process leads to the random percolation process. In Fig. \ref{fig4} we give a schematic of a lattice, where one can easily see that for two occupied sites (black) there are six sites with at least one occupied neighbor (red) in the attraction model (left panel), while there are 8 of them in the repulsion one (right panel). As a result of this, for a given density of occupied sites $p$, there is higher probability of having random percolation in repulsion model than in the attraction one. In higher densities $(0.5 < p < 1)$ the system undergoes a phase transition and the majority of sites are occupied. At this last state there are only few sites that are not occupied and the number of those which are isolated (no nearest neighbors) becomes even smaller. Thus, almost each new randomly chosen site is directly occupied in the attraction model because newly added sites are attracted by occupied neighbors. But in the repulsion model, since the number of isolated sites is almost zero, random percolation occurs again (because the probability to choose an isolated site is very small). Thus, at this point, repulsion and attraction models are both equivalent to the well-known random percolation. The simulations (Fig. \ref{fig1}) and the qualitative approach (Fig. \ref{fig4}) show that the rule for the attraction model and the one for the repulsion are not exactly equivalent during the percolation transition process. This is the reason why the curves for these two models are not symmetric to the one of classical random percolation. This asymmetry still exists even for $k$ values higher than one (Fig. \ref{fig5}).

We obtain results with different $k$ values for the attraction and repulsion models. We illustrate the results for both models in the same plot, in Fig. \ref{fig5} which shows the critical threshold $p_c$ for both attraction and repulsion models and for different values of attempts $k$. We observe in Fig. \ref{fig5} that the asymmetry between the two models still exists even for larger $k$ values. In addition, we observe that there is a minimum $p_c$ for the attraction model for $k \approx 15$. For $k>15$ the speed up for the attraction model does not have any further effect and further increase of the number of attempts $k$ results again to the delay of the critical point. The shape of the curve in Fig. \ref{fig5} with the minimum, giving the delay is quite unexpected, but it can be explained by the fact that after this minimum value the newly added sites are attracted to already existing clusters, with hardly any new clusters being formed. No new clusters appear (or the rate of appearance is limited). This is similar to crystallization starting from active centers. Thus, the formation of the infinite cluster which is spanning the entire system is delayed because there is a small number of clusters which are growing simultaneously and need more time to grow in all directions until they touch each other to form a larger one. 
In the limit of $k \rightarrow \infty$ there is only one nucleus that grows. The larger the $k$ value, the stronger is the similarity of the model to that of overall crystallization kinetics \cite{avrami}. When a large number of attempts $k$ is made most of the newly deposited sites are attached to the already existing clusters (except for the initial stages). In a sense at this range the filling of the lattice is equivalent to a crystallization process which proceeds from an initially small number of active centers.  

\section{Discussion and Conclusions}
In this work we have simulated two different models of the percolation phase transition, which depend on the method used to fill the lattice sites. The impetus for this work has been the fact that one can vary the location of the critical point almost at will by varying the values of the building parameters of the clusters. This variation alone may be important when preparing a new system with custom-made properties, which can now be tailor-made according to the needed specifications. We simulated the direct and reverse cases of the well-known Achlioptas processes using both the product and the sum rule. The reverse Achlioptas process is the one in which the probe site maximizes the product (or the sum) of the sizes of the clusters that is about to connect, as opposed to minimizing these as in the original Achlioptas processes. We introduce two new models based in the control of the local environment of the site to be added. These new models are based on the occupancy of the nearest neighbors of the probing sites. The attraction model promotes the merging of sites to form larger clusters and the repulsion model, which is the reverse process, promotes the isolation of occupied sites. 
We located the exact position of the critical density for all seven models that we have examined. Note that the position of the critical point is the value of density $p$ at the inflection point of the curves given in Fig. \ref{fig1} or, in a better way, the density of the peak of the first derivative of percolation strength $\frac{dP_{max}}{dp}$, as given in Fig. \ref{fig2}.
We compared the new findings with well known results, and we found excellent agreement. Our results show that the two new models belong to the same universality class as the classical percolation transition. 
Also, we extended the newly introduced models by further promoting the merging or the isolating of clusters using a parameter which controls random percolation at attraction and repulsion models.  
We explained the asymmetry that appears in the attraction and repulsion model around the normal critical transition point. Our results make it possible to control the location of the critical point by controlling the method of the preparation of the system. In principle, one could use fractional, non-integer values of $k$ (as done in Ref. \cite{paris2013}) and, thus, produce a new system with the exactly desired critical transition point.     
\begin{acknowledgments}
We thank S. Havlin for useful suggestions and discussions. This work used the European Grid Infrastructure(EGI) through the National Grid Infrastructures NGI\_GRNET, HellasGRID as part of the SEE Virtual Organization and was supported by the EC-funded FP7 project MULTIPLEX Grand number 317532.
\end{acknowledgments}

\bibliography{references}

\end{document}